\documentclass{article}
\usepackage{arxiv}
\usepackage[utf8]{inputenc} 
\usepackage{hyperref}       
\usepackage{amsfonts}       
\usepackage{amsmath}
\usepackage{color}
\usepackage{graphicx}
\usepackage{subfigure}
\usepackage{cite}
\usepackage{braket}
\usepackage{statmath}
\usepackage{float}
\usepackage[title]{appendix} 

\title{Spectral analysis of Dirac materials in position-dependent magnetic and electric fields via Heun functions.}
\author{Daniel O-Campa$^{\dagger,1}$, Omar Pedraza$^{*,2}$, L. A. López$^{*,3}$ and Erik Díaz-Bautista$^{\dagger,4}$\\
$^{\dagger}$Escuela Superior de Física y Matemáticas del Instituto Politécnico Nacional, 07738 Mexico City, Mexico.\\
$^{*}$\'Area Acad\'emica de Matem\'aticas y F\'isica, UAEH, 
	Carretera Pachuca-Tulancingo Km. 4.5,\\ C. P. 42184, Mineral de la Reforma, Hidalgo, M\'exico.\\
e-mail:  dortizca@ipn.mx$^1$, omarp@uaeh.edu.mx$^2$, lalopez@uaeh.edu.mx$^3$, ediazba@ipn.mx$^4$}
\begin{document}
\maketitle
\begin{abstract}
This work focuses on the study of the spectral problem for Dirac materials immersed in position-dependent magnetic and electric fields. To achieve this, the system of differential equations satisfied by the eigenfunction components of the Hamiltonian has been decoupled, and the solutions for some specific cases have been analyzed using Heun functions, which provide us with a quantization relation and allow us to determine the solutions for the bound states.
\end{abstract}
\section{Introduction}
\label{1}
In recent years, Dirac materials have captured the attention of numerous researchers, not only in condensed matter physics but also in other branches of physics. This interest stems from the fact that equations that describe their charge carriers in a low energy regime (near the Fermi level) can be used to analyze analogous physical systems or to explore the possibility of finding exact solutions applicable to various types of interactions \cite{Feng2016,Feng20161,LopezBezanilla2016,Zhao2018}.

For example, a few years ago, supersymmetric quantum mechanics (SUSY QM) -- a well-established formalism for analytically solving physical systems described by Schrödinger-like differential equations -- was successfully applied to the study of interactions between monolayer and bilayer graphene electrons and perpendicular magnetic fields \cite{Kuru2009,concha2018,Fernández_C_2020}. This approach enabled the derivation of exact solutions for configurations involving inhomogeneous magnetic fields, as the resulting equations exhibit a direct connection with the factorization method employed in SUSY QM. However, additional interactions, such as external electric fields, have since been incorporated into the physical models of graphene, making the analysis through exact solutions considerably more challenging \cite{Sari2015,Islam2018,Jafari2019}. Furthermore, the study of graphene has opened the door to the research of other two-dimensional Dirac materials, which have emerged as promising candidates for technological applications in fields such as valleytronics and spintronics \cite{nwu080,Kundu2016,Schaibley2016,Ang2017}. Consequently, the interaction of these materials with external electric and magnetic fields has regained attention from a mathematical physics perspective.

In this work, we revisit the problem of anisotropic Dirac materials with tilted cones, focusing on the interaction of their charge carriers with position-dependent external magnetic and electric fields. We aim to successfully implement a direct approach in which the energy spectrum and its corresponding eigenfunctions can be determined when more general configurations than those previously reported in the literature are used \cite{celeita2020,Mojica-Zárate_2023,campa2024, O-Campa_2024, Ateş_2023}.

The organization of this paper is as follows. In Section \ref{2}, we present the effective Hamiltonian to be analyzed and an appropriate way to decouple the set of equations associated with the eigenvalue problem is discussed. Section \ref{3} provides  a list of solvable cases using the developed method, along with the polynomial solutions of the Heun equations. Finally, we conclude with remarks on the results obtained in Section \ref{4}.
\section{The eigenvalue problem}
\label{2}
At low energies, a wide variety of two-dimensional materials exhibit linear or quasilinear dispersion relations due to their lattice structure, which also induces anisotropies or Dirac cones tilting. For this reason, they are often referred to as Dirac materials, as their dynamics are described by a Dirac-like Hamiltonian given by
\begin{equation}
\mathcal{H}_0 = \nu \left[v_x p_x \sigma_x + v_y p_y \sigma_y + v_{\rm t} p_y \sigma_0 \right],
\label{eq:1}
\end{equation}
where $\nu = \pm 1$ represents the valley index, $v_x, v_y$ are the anisotropic Fermi velocities, $v_{\rm t}$ is the velocity term arising from the Dirac cone tilt, $\sigma_{x}, \sigma_{y}$ are the Pauli matrices, and $\sigma_0$ is the $2 \times 2$ identity matrix. When the system is affected by the presence of an external electric field $\mathbf{E} = \mathcal{E}(x)\hat{e}_x$ in the plane of the material (the $x$-$y$ plane) and a magnetic field $\mathbf{B} = \mathcal{B}(x)\hat{e}_z$ perpendicular to it, both with translational symmetry with respect to the $y$ coordinate, the Hamiltonian in Equation \eqref{eq:1} must be modified to include the information of these fields through the minimal coupling rule as
\begin{equation}
\mathcal{H} = \nu \left[ v_x p_x \sigma_x + v_y \left(p_y + \mathcal{A}_y(x)\right)\sigma_y + v_{\rm t} \left(p_y + \mathcal{A}_y(x)\right) \sigma_0 \right] - \phi(x)\sigma_0,
\label{eq:2}
\end{equation}
where $\phi(x)$ is the scalar potential that generates the electric field and $\mathcal{A}_y$ is the component of the vector potential in the Landau gauge ($\mathbf{A} = \mathcal{A}_y(x)\hat{e}_y$) that generates the magnetic field. Due to the translational invariance in the $y$-direction, the Hamiltonian given in Equation \eqref{eq:2} commutes with the operator $p_y$; therefore, it is possible to express its eigenfunctions following the ansatz:
\begin{equation}
\Psi(x, y) = e^{ik_y y} \bar{\Psi}(x),
\label{eq:3}
\end{equation}
with $\bar{\Psi}(x) = \left( \psi^+(x), i \psi^-(x) \right)^{\rm T}$. Thus, the eigenvalue problem $\mathcal{H} \Psi(x, y) = E \Psi(x, y)$ leads us to the following differential equation for $\bar{\Psi}(x)$:
\begin{equation}
\left[-iv_x \frac{\rm d}{{\rm d}x} \sigma_x + v_y \left(k_y + \mathcal{A}_y(x)\right)\sigma_y + \left(v_{\rm t} k_y + v_{\rm t} \mathcal{A}_y(x) - \nu \phi(x) - \nu E \right)\sigma_0 \right] \bar{\Psi}(x) = 0.
\label{eq:4}
\end{equation}
To determine the set of eigenfunctions and eigenvalues of the Hamiltonian, it is necessary to find the solutions corresponding to the previous equation. However, this leads us to a system of coupled differential equations for the components $\psi^{\pm}(x)$ of the spinor $\bar{\Psi}(x)$. Hence, in the next section, we will present an approach that allows us to find the solutions exactly.
\subsection{Decoupling the differential system}
In general, the solutions to Equation \eqref{eq:4} strongly depend on the form of the fields. Therefore, we proceed as follows. First,  by multiplying Equation \eqref{eq:4} by $i \sigma_x$ on the left-hand side and algebraically manipulating, the system presented below is obtained:
\begin{equation}
\frac{\rm d}{{\rm d}x} \bar{\Psi}(x) = \left[i {\rm F_1}(x) \sigma_x + {\rm F_2}(x) \sigma_z \right] \bar{\Psi}(x), 
\label{eq:5}
\end{equation}
where
\begin{equation}
{\rm F}_1(x) = \nu \frac{E - \nu v_{\rm t} k_y + \phi(x) - \nu v_{\rm t} \mathcal{A}(x)}{v_x}, \quad {\rm F}_2(x) = \frac{v_y}{v_x} \left(k_y + \mathcal{A}_y(x)\right). 
\label{eq:6}
\end{equation}
The equation above leads us to a set of coupled differential equations for the components $\psi^{\pm}(x)$, given by:
\begin{equation}
\left(\mp \frac{\rm d}{{\rm d}x} + {\rm F}_2(x) \right) \psi^{\pm}(x) = {\rm F}_1(x) \psi^{\mp}(x).
\label{eq:7}
\end{equation}
If we now define the operators $\mathcal{L}^+$ and $\mathcal{L}^-$ as:
\begin{equation}
\mathcal{L}^{\pm} = \left(\mp \frac{1}{{\rm F}_1(x)} \frac{\rm d}{{\rm d}x} + \frac{{\rm F}_2(x)}{{\rm F}_1(x)} \right),
\label{eq:8}
\end{equation}
then the system of equations in \eqref{eq:7} can be rewritten as:
\begin{equation}
\mathcal{L}^{\pm} \psi^{\pm}(x) = \psi^{\mp}(x).
\label{eq:9}
\end{equation}
Therefore, by applying the operator $\mathcal{L}^{\mp}$ to both sides of Equation \eqref{eq:9}, the system of equations decouples, and after some algebra, two decoupled second-order homogeneous linear differential equations for the components $\psi^{\pm}(x)$ are obtained, which turn out to be:
\begin{equation}
\left[- \frac{\rm d^2}{{\rm d}x^2} + \frac{{\rm F}_1^{'}(x)}{{\rm F}_1(x)} \frac{\rm d}{{\rm d}x} + \left( {\rm F}_2^2(x) - {\rm F}_1^2(x) \pm \frac{{\rm F}_1(x) {\rm F}_2^{'}(x) - {\rm F}_1^{'}(x) {\rm F}_2(x)}{{\rm F}_1(x)} \right) \right] \psi^{\pm}(x) = 0.
\label{eq:10}
\end{equation}
Note that Equation \eqref{eq:10} represents a decoupled system of differential equations, which now allows us to directly find the components of the spinor in Equation \eqref{eq:3} for arbitrary field profiles. However, in general, this system is not composed of eigenvalue equations that allow for the direct identification of the energy. Nevertheless, it is possible to obtain exact solutions for certain field profiles, as shown below.
\section{Examples of exact solvability}
\label{3}
In the following, we analyze some examples of external electric and magnetic field profiles for which it is possible to determine exact solutions of the system of equations in ~\eqref{eq:10}, and thus obtain the energy spectrum and the corresponding eigenfunctions. To this end, we begin by assuming that the magnetic and electric field profiles are proportional to each other, that is, they differ by a multiplicative constant denoted by $-v_{\rm d}$, which is read as
\begin{equation}
\phi(x)=-v_{\rm d}\mathcal{A}_y(x).
\label{eq:11}
\end{equation}
It can be observed that for $v_{\rm d} = 0$, the electric field is zero. If $v_{\rm d} > 0$, then both fields point in the same direction (either positive or negative), and if $v_{\rm d} < 0$, then one field points in the positive direction while the other points in the negative direction. All three of these cases can be treated without loss of generality by assuming that $v_{\rm d}>0$ and subsequently taking the appropriate limit. For this reason, we will consider in the following discussion that both fields point in the positive direction.\\
\\
Now, to simplify the expressions and calculations, we first define the following quantities:
\begin{equation}
\beta_{\nu}=\frac{v_{\rm d}+\nu v_{\rm t}}{v_y},\quad\kappa=\frac{kv_y+\beta_{\nu}\left(E-\nu  v_{\rm t}k\right)}{\sqrt{1-\beta_{\nu}^2} v_x},\quad
\epsilon=\left(\frac{E+v_{\rm d}k}{\sqrt{1-\beta_{\nu}^2} v_x}\right)^2,\quad{\rm where}\quad|\beta_{\nu}|<1.
\label{eq:12}
\end{equation}
With these elements, we now proceed to analyze three examples of field profiles.
\subsection{Constant fields}
As a first example, we will consider the standard case on which both the magnetic field and the electric field are constant, namely, $\mathbf{B}=\mathcal{B}_0\hat{e}_z$ and $\mathbf{E}=\mathcal{E}_0\hat{e}_x$ with $\mathcal{B}_0,\mathcal{E}_0>0$ (in this way, $v_{\rm d}=\mathcal{E}_0 / \mathcal{B}_0$). Then, the $ y $-component of the vector potential can be chosen as $ \mathcal{A}_y(x)=\mathcal{B}_0x $ and $\phi(x)=-\mathcal{E}_0x$. Furthermore, we define the following quantities:
\begin{equation}
\Omega=2\frac{v_y}{v_x}\sqrt{1-\beta^2_{\nu}}\mathcal{B}_0,\quad
\bar{\alpha}=-2,\quad\bar{\beta}=\frac{2}{\beta_{\nu}}\sqrt{\frac{2\epsilon}{\Omega}}
,\quad\bar{\gamma}=\frac{2\epsilon}{\Omega},\quad\bar{\delta}^{\pm}=\mp\frac{2}{\beta_{\nu}}\sqrt{\frac{2(1-\beta^2_{\nu})\epsilon}{\Omega}},
\label{eq:13}
\end{equation}
and applying the change of variable given by
\begin{equation}
z=\sqrt{\frac{\Omega}{2}}\left[x+2\sqrt{1-\beta_{\nu} ^2}\frac{\nu\, v_{\rm t}k-E}{ v_x\beta_{\nu}\Omega}\right],
\label{eq:14}
\end{equation}
Equation \eqref{eq:10} transforms into
\begin{equation}
\left\lbrace-\frac{\rm d^2}{{\rm d}z^2}+
\frac{1}{z}
\frac{\rm d}{{\rm d}z}+\left[\left(z+\frac{\bar{\beta}}{2}\right)^2-\bar{\gamma}+\frac{\bar{\delta}^{\pm}}{2z}\right]
\right\rbrace\phi^{\pm}(z)=0,
\label{eq:15}        
\end{equation}
where $\psi^{\pm}(x)=\phi^{\pm}(z)$. It can be observed that asymptotically, Equation \eqref{eq:15} behaves like the one corresponding to a harmonic oscillator, which makes it natural to propose the following structure for $\phi^{\pm}$:
\begin{equation}
\phi^{\pm}(z)=e^{-\frac{(z+\frac{\bar{\beta}}{2})^2}{2}}f^{\pm}(z),
\label{eq:16}    
\end{equation}
being $f^{\pm}$ functions to be determined while the Gaussian form guarantees that $\phi^{\pm}$ are asymptotically correct solutions. By substituting \eqref{eq:16} into \eqref{eq:15}, the following differential equation for $f^{\pm}(z)$ is obtained
\begin{equation}
\left\lbrace\frac{{\rm d}^2}{{\rm d}z^2}+\left(\frac{1+\bar{\alpha}}{z}-\bar{\beta}-2z\right)\frac{{\rm d}}{{\rm d}z}+\left(\bar{\gamma}-\bar{\alpha}-2-\frac{\bar{\delta}^{\pm}+\bar{\beta}\left(1+\bar{\alpha}\right)}{2z}\right)\right\rbrace f^{\pm}(z)=0.
\label{eq:17}
\end{equation}
In the literature, Equation \eqref{eq:15} is commonly referred to as the biconfluent Heun differential equation\cite{ARRIOLA1991161,ISHKHANYAN201779}, thus, $f^{\pm}(z)=\text{HeunB}(\bar{\alpha},\bar{\beta},\bar{\gamma}, \bar{\delta}^{\pm};z)$. However, since the function $ \psi^{\pm}$ must be square integrable, it is essential that $ f^{\pm}(z) $ be a polynomial of finite order $n$, a condition that is satisfied if and only if 
\begin{equation}
\bar{\gamma}-\bar{\alpha}-2=2n.
\label{eq:18}
\end{equation}
We must emphasize that even though the differential equations in \eqref{eq:10} satisfied by $\psi^{\pm}$ are not eigenvalue equations with energy as the eigenvalue. Actually, Equation \eqref{eq:18} acts as a quantization condition that allows us to determine the energy spectrum of the system. Thus, from equations \eqref{eq:12} and \eqref{eq:13}, it follows that
\begin{equation}
E_n = -k\,v_{\rm d}+\mu v_x \left(1 - \beta_{\nu}^2 \right)^{\frac{1}{2}} \sqrt{n\,\Omega},\quad
n=0,1,....
\label{eq:19}
\end{equation}
being $\mu=\pm1$, where the plus (minus) sign corresponds to the conduction (valence) band. Now, based on the information provided by the quantization relation in \eqref{eq:18} and to determine the set of eigenfunctions corresponding to \eqref{eq:19}, we will show that the solutions $f^{\pm}$ to the Heun equation in \eqref{eq:17} can be expressed as a linear combination of Hermite polynomials.
\subsubsection{Polynomial solutions of the biconfluent Heun differential equation}
By considering a finite-order polynomial $n$ as a solution to the biconfluent Heun differential equation in \eqref{eq:17}, the relation $\bar{\gamma}-\bar{\alpha}-2=2n$ is obtained. On the other hand, if we assume that such a polynomial solution can be written in terms of Hermite polynomials, then the following condition for $f^{\pm}$ must be satisfied:
\begin{equation}
f^{\pm}(z)=\sum_{k=0}^nA^{\pm}_kz^k=\sum_{k=0}^NC^{\pm}_k u_k(z),\quad
u_k(z)=\mbox{H}_{\tau_0+k}\left(s_0(z+z_0)\right),
\label{eq:20}
\end{equation}
where $C^{\pm}_k$ are the coefficients to be determined and $\tau_0$, $s_0$, $z_0$ are constants that we can choose at will as long as the convergence of the function and equality with the sum of the powers are guaranteed.\\
\\
In this regard, the Hermite polynomials in \eqref{eq:20} satisfy the following differential equation
\begin{equation}
u_k''(z) - 2s_0^2(z+z_0) u_k'(z) + 2s_0^2\tau_k u_k(z) = 0,\quad\tau_k=\tau_0+k,    
\label{eq:21}
\end{equation}
in addition to the recurrence relations:
\begin{equation}
u_k'(z)=2s_0\tau_k u_{k-1}(z),\quad
s_0(z+z_0)u_k(z)=\tau_{k}u_{k-1}(z)+\frac{1}{2}u_{k+1}(z).
\label{eq:22}
\end{equation}
Then, by substituting $f^{\pm}$ from \eqref{eq:20} into Equation \eqref{eq:17} and using \eqref{eq:21} to replace the second-order derivative of $u_k$, it follows that
\begin{equation}
\sum_{k=0}^NC^{\pm}_k\left\lbrace\left[\left(2s_0^2-2\right)z^2+\left(2s_0^2z_0-\bar{\beta}\right)z+1+\bar{\alpha}\right]u'_k(z)+\left[\left(\bar{\gamma}-\bar{\alpha}-2-2s_0^2\tau_k\right)z-\frac{\bar{\delta}^{\pm}+\bar{\beta}(1+\bar{\alpha})}{2}\right]u_k(z)\right\rbrace=0.
\label{eq:23}
\end{equation}
In this manner, we can use the freedom we have over $s_0$ and $z_0$ to eliminate the terms proportional to $z^2u'_k$, and $zu'_k$ by choosing
\begin{equation}
s_0=\pm1,\quad z_0=\frac{\bar{\beta}}{2}.
\label{eq:24}
\end{equation}
Thus, by making use of the relationship between $\bar{\gamma}$, $\bar{\alpha}$, and $n$, as well as equations in \eqref{eq:24} and the recurrence relations in \eqref{eq:22}, Equation \eqref{eq:23} leads us to the following equality:
\begin{equation}
\sum_{k=0}^NC^{\pm}_k\left[R_ku_{k+1}(z)+Q^{\pm}_ku_{k}(z)+P_ku_{k-1}(z)\right]=0,
\label{eq:25}
\end{equation}
being
\begin{equation}
R_k=\left(n-\tau_k\right),\quad
Q^{\pm}_k=s_0\bar{\beta}\left(\tau_k-n+\frac{1\pm\left(1-\beta_{\nu}^2\right)^{\frac{1}{2}}}{2}\right),\quad
P_k=2(n-\tau_k-1)\tau_k.
\label{eq:26}
\end{equation}
By considering that the Hermite polynomials constitute a complete basis, the following conditions must be satisfied simultaneously:
\begin{equation}
P_0=0\quad\mbox{and}\quad R_N=0,
\label{eq:27}
\end{equation}
which allows to determine the values of $\tau_0$ and $N$, since the first one needs that
\begin{equation}
\tau_0=0\quad\mbox{or}\quad\tau_0=n-1,
\label{eq:28}
\end{equation}
while the second one holds if
\begin{equation}
\tau_0=n-N.
\label{eq:29}
\end{equation}
Equations \eqref{eq:28} and \eqref{eq:29} imply that $N=n$ or $N=1$. Although both choices lead to the same solution expressed in two different bases, the first one involves a more complicated calculation due to the following factors: 1) the upper limit of the sum depends on the degree of the polynomial considered for the solution; 2) regardless of the value of $n$, $P_{n-1}=0$, whereas with the other choice $P_1=0$ only occurs if $n=0$. Therefore, determining the coefficients $C^{\pm}_k$ is more laborious. For this reason, we will consider only the case $N = 1$, then, it turns out that
\begin{equation}
\tau_0=n-1,\quad
R_k=\left(1-k\right),\quad
Q^{\pm}_k= s_0\frac{\bar{\beta}}{2}\left(2k-1\pm\left(1-\beta_{\nu}^2\right)^{\frac{1}{2}}\right),\quad
P_k=2k(1-n-k).
\label{eq:30}
\end{equation}
Thus, taking into account that $ u_k(z) $ form a basis, Equation \eqref{eq:25} can be expressed as:
\begin{equation}
\mathbf{M} \vec{C^{\pm}} = \vec{0}, \quad \text{with } 
\mathbf{M} = \begin{pmatrix}
R_0 & Q^{\pm}_1 \\
Q^{\pm}_0 & P_1
\end{pmatrix}, \quad
\vec{C^{\pm}} = \begin{pmatrix}
C^{\pm}_0 \\
C^{\pm}_1
\end{pmatrix}.
\label{eq:31}
\end{equation}
This system has a non-trivial solution if and only if $ \mathbf{M} $ is non-invertible, i.e., its determinant vanishes:
\begin{equation}
\det(\mathbf{M}) = R_0 P_1 - Q^{\pm}_1 Q^{\pm}_0 = 0.
\label{eq:32}
\end{equation}
Then, by considering the expressions in \eqref{eq:30} and manipulating algebraically, it can be shown that indeed $\det(\mathbf{M})=0$. Hence, $C^{\pm}_0$ and $C^{\pm}_1$ are not independent but satisfy the following relation:
\begin{equation}
C^{\pm}_1Q^{\pm}_1+C^{\pm}_0 R_0=C^{\pm}_1s_0\frac{\bar{\beta}}{2}\left(1\pm(1-\beta_{\nu}^2)^{\frac{1}{2}}\right)+C^{\pm}_0=0.
\label{eq:33}
\end{equation}
However, due to the quantization condition, it follows that $\bar{\beta}\propto \sqrt{n}$, which leads to the following two cases:
\begin{itemize}
\item[1] \textbf{Case $n=0$.} For this case, $\bar{\beta} = 0$, thus $C^{\pm}_0 = 0$. In this way, the solution is given by
\begin{equation}
f^{\pm}(z) = C^{\pm}_1 \mbox{H}_0(z + z_0),
\label{eq:34}    
\end{equation}
where $C^{\pm}_1$ will be determined by the normalization of the final solution, and the argument was chosen simply as $z + z_0$, since for both possible values of $s_0$, the Hermite polynomial $\mbox{H}_0$ is equal to 1.
\item[2] \textbf{Case $n\geq1$}. For this case, $\bar{\beta}\neq 0$. Thus, according to Equation \eqref{eq:33}, it must hold that the solutions $f^{\pm}$ to the biconfluent Heun differential equation are given by:
\begin{equation}
f^{\pm}(z) = C^{\pm}_1 \left(-s_0\frac{\bar{\beta}}{2}\left(1\pm(1-\beta_{\nu}^2)^{\frac{1}{2}}\right)\mbox{H}_{n-1} \left( s_0 \left( z + z_0 \right) \right) + \mbox{H}_{n} \left( s_0 \left(z +z_0 \right) \right) \right).
\label{eq:35}
\end{equation}
Notice that, regardless of the choice of $s_0 = \pm 1$, both values will only generate solutions with a global phase difference. Therefore, without loss of generality, we can choose $s_0 = 1$. 
\end{itemize}
Finally, since $ z $ and $ z_0 $ depend implicitly on the energy, and therefore on $ n $, it is convenient to define the variable
\begin{equation}
\theta_n = z + z_0 = \sqrt{\frac{\Omega}{2}} \left( x + 2\frac{v_y}{v_x}\frac{k}{\sqrt{1 - \beta_{\nu}^2}\,\Omega} \right)+\mu\beta_{\nu} \sqrt{2n}.
\label{eq:36}
\end{equation}
In this manner, by taking into account equations \eqref{eq:16}, \eqref{eq:34}, and \eqref{eq:35}, 
the normalized eigenfunctions corresponding to the energies of Equation \eqref{eq:19} are:
\begin{equation}
\Psi_n(x,y) = \frac{e^{ik_y y}}{\sqrt{2^{1- \delta_{n,0}}}} \begin{pmatrix} \psi^+_n(x) \\ \\ i\nu\mu\psi^-_n(x) \end{pmatrix},
\label{eq:37}
\end{equation}
where $\psi^{\pm}_n$ can be written as in \cite{celeita2020,campa2024,Lukose2007}:
\begin{align}
\psi^+_n(x)&=\mathcal{N}_n \exp\left(-\frac{\theta_n^2}{2}\right) \left(\left( 1 + \left( 1 - \beta_{\nu}^2 \right)^{\frac{1}{2}} \right)^{\frac{1}{2}} \left( 1 - \delta_{n,0} \right) \sqrt{2n} \mbox{H}_{n-1} \left( \theta_n \right)-\frac{\mu \beta_{\nu}}{\left( 1 + \left( 1 - \beta_{\nu}^2 \right)^{\frac{1}{2}} \right)^{\frac{1}{2}}} \mbox{H}_{n} \left( \theta_n \right)\right),
\label{eq:38}\\
\psi^-_n(x)&=\mathcal{N}_n\mbox{exp}\left(-\frac{\theta_n^2}{2}\right)
\left(- \frac{\mu \beta_{\nu} \sqrt{2n}}{\left( 1 + \left( 1 - \beta_{\nu}^2 \right)^{\frac{1}{2}} \right)^{\frac{1}{2}}} \left( 1 - \delta_{n,0} \right) \mbox{H}_{n-1} \left(\theta_n\right)+\left( 1 + \left( 1 - \beta_{\nu}^2 \right)^{\frac{1}{2}} \right)^{\frac{1}{2}} \mbox{H}_{n} \left(\theta_n \right)\right),
\label{eq:39}
\end{align}
with $\mathcal{N}_n = \frac{1}{\sqrt{2^n n!}} \left( \frac{\Omega}{8\pi} \right)^{\frac{1}{4}}$. Thus, knowing that the probability density and probability current for the Hamiltonian in Equation \eqref{eq:1} are given by
\begin{equation}
\rho(x,y)=\Psi^{\dagger}(x,y)\Psi(x,y),\quad
\vec{\mathcal{J}}(x,y)=\Psi^{\dagger}(x,y)\vec{j}\Psi(x,y),\quad
j_x=\nu v_x\sigma_x,\quad
j_y=\nu\left(v_y\sigma_x+v_{\rm t}\sigma_0\right),
\label{corrientes}
\end{equation}
for the excited states we also have
\begin{equation}
\rho_n(x,y)=\rho_n(x)=\bar{\Psi}^{\dagger}_n(x)\bar{\Psi}_n(x),\quad
\mathcal{J}_{x,n}=0,\quad
\mathcal{J}_{y,n}=2^{\delta_{n,0}}\mu v_y\psi_n^+(x)\psi_n^-(x)+\nu\rho_n(x).\quad
\label{CorrrientesN}
\end{equation}
These expressions are also valid for the following cases. A brief discussion of these results is presented below.
\subsubsection{Discussion}
\begin{figure}[h!] 
\begin{center}
\includegraphics[width=0.55\textwidth]{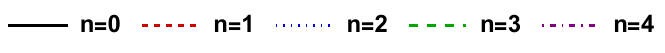}
\subfigure[]{\includegraphics[width=8cm, height=5.7cm]{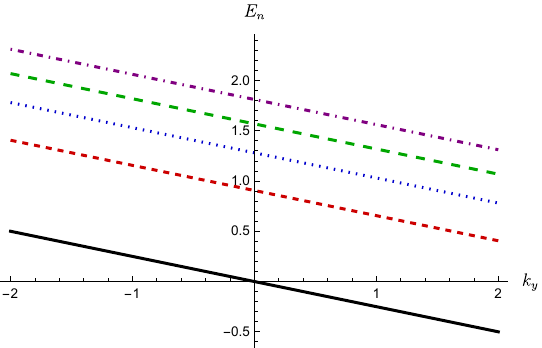}}
\subfigure[]{\includegraphics[width=8cm, height=5.7cm]{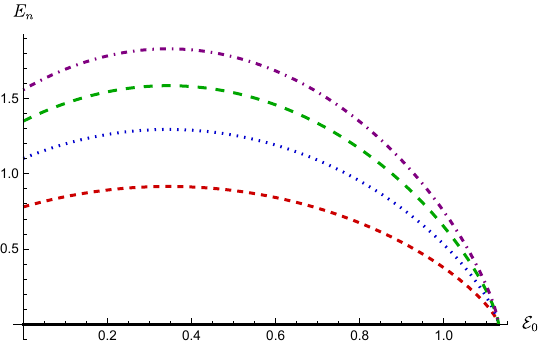}}
\subfigure[]{\includegraphics[width=8cm, height=5.7cm]{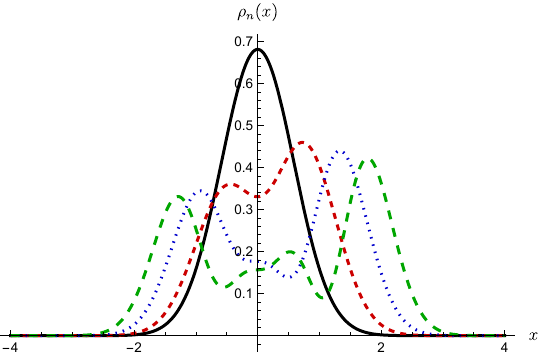}}
\subfigure[]{\includegraphics[width=8cm, height=5.7cm]{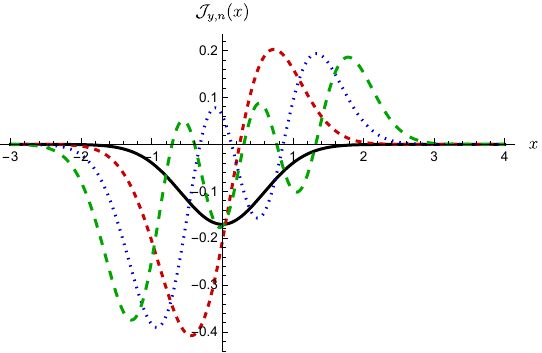}}
\caption{(a) Plots of the energy spectrum in \eqref{eq:19} as a function of the wavenumber $k_{y}$ and (b) as a function of the electric field strength $\mathcal{E}_{0}$. (c) Plot of the probability density $\rho_n(x)$ and (d) $y$-component current density $\mathcal{J}_{y,n}(x)$ corresponding to the eigenfunctions in \eqref{eq:37}. The values have been set as $k_{y}=0$, $\mathcal{B}_0=\nu=\mu=1$, $\left\lbrace v_x,\;v_y,\;v_{\rm t},\;v_{\rm d}\right\rbrace=\left\lbrace 0.534,\;0.785,\;-0.345,\;0.25\right\rbrace$.}
\label{F1}
\end{center}
\end{figure}

For the given configuration of electric and magnetic field profiles, the energy spectrum described in Equation \eqref{eq:19} is discrete, infinite, and dispersive, exhibiting a linear behavior with respect to the wave number in the $y$-direction. This linearity presents a negative slope corresponding to $-v_{\rm d}$, i.e., the negative ratio of the field amplitudes (see Figure \ref{F1}a). For that reason, if the electric field is null ($v_{\rm d} = 0$), the energy levels become completely flat with respect to $k$ (non-dispersive), regardless of the anisotropic velocities or the value of $v_{\rm t}$. On the other hand, it is worth noting that as the parameter $\beta_\nu$ approaches $1$ (or $-1$)—that is, when the electric field amplitude approaches the critical value $\mathcal{E}_0 = \mathcal{B}_0 (v_y - \nu v_{\rm t})$—the energy levels tend to become degenerate at a single value of $-v_{\rm d} k$ (see Figure~\ref{F1}b).

Finally, while it is evident that the probability density depends on the valley and band indices (see Figure \ref{F1} c), a more interesting result arises from the $y$-component of the probability current. Due to the tilt of the Dirac cones, a non-zero current is generated in the ground state regardless of the values of the anisotropic velocities $v_x$ and $v_y$ (see Figure~\ref{F1} d), in contrast to materials such as graphene\cite{Kuru2009}.
\subsection{Exponential decaying fields}
The second example we consider involves both fields exhibiting an exponentially decaying profile, namely, \mbox{$\mathbf{B} = \mathcal{B}_0 e^{-\alpha x} \hat{e}_z$} and $\mathbf{E} = \mathcal{E}_0 e^{-\alpha x} \hat{e}_x$, where $\mathcal{B}_0, \mathcal{E}_0, \alpha > 0$. In this configuration $v_{\rm d} = \mathcal{E}_0 / \mathcal{B}_0$, as before. The corresponding potentials can be chosen as $\mathcal{A}_y(x) =-\frac{\mathcal{B}_0}{\alpha} e^{-\alpha x}$ and $\phi(x) = \frac{\mathcal{E}_0}{\alpha} e^{-\alpha x}$. Moreover, if we define the following quantities:
\begin{align}
\mathcal{F}&=\sqrt{1-\beta_{\nu}^2}\frac{v_y}{v_x}\frac{\mathcal{B}_0}{\alpha},\quad
\bar{\alpha}=\frac{2 \left(\sqrt{\epsilon }-\beta_{\nu}  \kappa \right)}{\alpha  \beta_{\nu}},\quad
\bar{\beta}=\frac{2 \sqrt{\kappa ^2-\epsilon }}{\alpha },\quad\bar{\gamma}=-2,\nonumber\\
\bar{\delta}&=\frac{2 \kappa  \left(\beta_{\nu}\kappa -\sqrt{\epsilon }\right)}{\alpha ^2 \beta_{\nu} }\quad
\bar{\eta}^{\pm}=\frac{\alpha^2\beta_{\nu} \mp\alpha\sqrt{1-\beta_{\nu}^2} \sqrt{\epsilon }+2 \kappa \left(\sqrt{\epsilon }-\beta_{\nu}\kappa \right)}{\alpha^2\beta_{\nu}},    
\label{eq:40}
\end{align}
and considering the change of variable
\begin{equation}
z=\frac{\mathcal{F}\beta_{\nu} }{\beta_{\nu}\,\kappa - \sqrt{\epsilon}}e^{-\alpha x},
\label{eq:41}
\end{equation}
then the set of differential equations in \eqref{eq:10} is transformed into
\begin{equation}
\left\lbrace\frac{\rm d^2}{{\rm d}z^2}-\frac{1}{z(z-1)}
\frac{\rm d}{{\rm d}z}+\left(
\frac{\bar{\delta}+\bar{\eta}-1}{z-1}-\frac{\bar{\beta}^2}{4 z^2}-\frac{\bar{\eta}-1}{z}
-\frac{\bar{\alpha}^2}{4}
\right)
\right\rbrace\phi^{\pm}(z)=0,
\label{eq:42}        
\end{equation}
where $\psi^{\pm}(x) = \phi^{\pm}(z)$. It is worth noting that, as \( x \to -\infty \), the previous equation behaves like a modified Helmholtz equation in one dimension.Therefore, the solution is asymptotically of the form $\phi^{\pm}(z)\propto e^{\frac{\bar{\alpha}}{2}z}$. Similarly, as \( x \to \infty \), the differential operator in Equation \eqref{eq:42} behaves like $\frac{{\rm d}^2}{{\rm d}z^2} + \frac{1}{z}\frac{{\rm d}}{{\rm d}z} -\frac{\bar{\beta}^2}{4 z^2}$, and hence the solution is asymptotically given by $\phi^{\pm}(z)\propto z^{\frac{\bar{\beta}}{2}}$. Based on these asymptotic behaviors, we propose that the solution takes the form
\begin{equation}
\phi^{\pm}(z)=z^{\frac{\bar{\beta}}{2}}
{\rm e}^{\frac{\bar{\alpha}}{2}z}f^{\pm}(z).
\label{eq:43}
\end{equation}
By substituting \eqref{eq:43} into \eqref{eq:42}, the following differential equation for $f^{\pm}(z)$ is obtained
\begin{equation}
\left[\frac{{\rm d}^2}{{\rm d}z^2}+\left(
\bar{\alpha}+\frac{\bar{\beta}+1}{z}+\frac{\bar{\gamma}+1}{z-1}\right)\frac{{\rm d}}{{\rm d}z}+
\left(\frac{\frac{\left(\bar{\beta}+1\right)\left(\bar{\alpha}-\bar{\gamma}-1\right)+1}{2}-\bar{\eta}^{\pm }}{z}+\frac{\frac{\left(\bar{\gamma}+1\right)\left(\bar{\alpha}+\bar{\beta}+1\right)-1}{2}+\bar{\delta}+\bar{\eta}^{\pm }}{z-1}\right)\right]f^{\pm}(z)=0.
\label{eq:44}
\end{equation}
Equation~\eqref{eq:44} is commonly referred as the confluent Heun differential equation\cite{MELIKDZHANIAN2021125037}, so that \mbox{$f^{\pm}(z)=\text{HeunC}(\bar{\alpha},\bar{\beta},\bar{\gamma}, \bar{\delta},\bar{\eta}^{\pm};z)$.} As in the previous case, for $\psi^{\pm}$ to be square-integrable, we must require that $f^{\pm}(z)$ be a polynomial of finite degree $n$. This condition is satisfied if
\begin{equation}
\frac{\bar{\delta}}{\bar{\alpha}}+\frac{\bar{\beta}+\bar{\gamma}+2}{2}=-n.
\label{eq:45}
\end{equation}
Similarly to the first case, the previous equation acts as a quantization condition, allowing us to determine the energy values, which are found to be:
\begin{equation}
E_n=-v_{\rm d}\,k+\alpha\,v_x\, \beta_{\nu}\,\sqrt{1-\beta_{\nu}^2}\,n+\mu\,\sqrt{1-\beta_{\nu}^2}\,\sqrt{ v_y^2\,k^2-\left(v_y\,k-\alpha\,v_x\,\sqrt{1-\beta_{\nu}^2}\,n\right)^2},\quad
n=0,1,\dots,
\label{eq:46}
\end{equation}
where, once again, \( \mu \) denotes the band parameter. By assuming that \(f^{\pm}\) admits a polynomial solution, it can be shown that it is possible to express it in terms of known orthogonal polynomials. In this case, the associated Laguerre polynomials offer the best fit, as will be shown below.
\subsubsection{Polynomial solutions of the confluent Heun differential equation}
In order to demonstrate that the polynomial solution for \(f^{\pm}\) can be written as a linear combination of associated Laguerre polynomials, we begin by defining the following parameters:
\begin{equation}
\bar{\xi}^{\pm}=\frac{\left(\bar{\beta}+1\right) \left(\bar{\alpha}-\bar{\gamma}-1\right)+1}{2}-\bar{\eta}^{\pm } 
,\quad
\bar{\zeta}^{\pm}=\frac{\left(\bar{\gamma}+1\right) \left(\bar{\alpha} +\bar{\beta}+1\right)-1}{2}+\bar{\delta} +\bar{\eta}^{\pm }.
\label{eq:47}
\end{equation}
In terms of the parameters \(\bar{\xi}^{\pm}\) and \(\bar{\zeta}^{\pm}\), the quantization condition is given by $\bar{\xi}^{\pm}+\bar{\zeta}^{\pm}=-\bar{\alpha}n$. On the other hand, if we assume that the polynomial solution can be expressed in terms of associated Laguerre polynomials, then the following condition must hold:
\begin{equation}
f^{\pm}(z) = \sum_{k=0}^n A^{\pm}_k z^k = \sum_{k=0}^N C^{\pm}_k\, u_k(z), \quad
u_k(z) = \mathrm{L}_{\tau_0 + k}^{\rho_0 - k}\left( s_0(z + z_0) \right),
\label{eq:48}
\end{equation}
where $C^{\pm}_k$ are coefficients to be determined, while $\tau_0$, $\rho_0$, $s_0$, and $z_0$ are constants that can be freely chosen, provided they ensure the convergence of the function and consistency with the polynomial expression.\\
\\
Moreover, the differential equation for the Laguerre polynomials is well known, and therefore the following identity holds:
\begin{equation}
u_k''(z) + \left(\frac{\rho_{k-1}}{z+z_0}-s_0\right)u_k'(z) + \frac{s_0\tau_k}{z+z_0} u_k(y) = 0,\quad\tau_k=\tau_0+k,\quad\rho_k=\rho_0-k,
\label{eq:49}
\end{equation}
and also the recurrence relations
\begin{equation}
u_k'(z) =-s_0 u_{k-1}(z),\quad
(z+z_0)u_k'(z) = \tau_{k+1}u_{k+1}(z) + (s_0 (z+z_0) - \rho_k )\,u_k(z)\quad{\rm for}\quad
1\leq \tau_k.
\label{eq:50}
\end{equation}
Thus, by substituting the ansatz in Equation \eqref{eq:48} into the confluent Heun differential equation given in Equation \eqref{eq:44}, and replacing the second derivative of $u_k$ using the relation in Equation \eqref{eq:49}, we obtain—after some algebraic manipulations—that:
\begin{align}
\sum_{k=0}^NC_k^{\pm}\Bigg\lbrace&
\left[(\bar{\zeta}+\bar{\xi}-s_0\tau_k)z^2+(s_0\tau_k-\bar{\xi}+z_0(\bar{\zeta}+\bar{\xi}))z-\bar{\xi}z_0\right]u_k(z)\nonumber\\
&+\left[ (\bar{\alpha} + s_0) z^3 
+ \left( \bar{\beta} + \bar{\gamma}+2 - \rho_{k-1}+ (s_0+\bar{\alpha})(z_0 - 1)\right) z^2\right.\nonumber\\
&+\left.\quad\left( -\bar{\beta} + \rho_{k-1} 
+ z_0 (-\bar{\alpha} + \bar{\beta} + \bar{\gamma} - s_0 + 2) - 1 \right) z 
- (\bar{\beta} + 1) z_0 \right] u_k'(z) 
\Bigg\rbrace=0.
\label{eq:51}
\end{align}
By exploiting the freedom in the choice of parameters, we can set $s_0$ and $z_0$ in such a way that the terms proportional to $z^3u'_k(z)$ and $u'_k(z)$ vanish. Thus, we fix:
\begin{equation}
s_0=-\bar{\alpha}, \quad
z_0 = 0.
\label{eq:52}
\end{equation}
With such a selection, a straightforward algebraic manipulation shows that Equation \eqref{eq:51} becomes equivalent to:
\begin{equation}
\sum_{k=0}^NC_k^{\pm}\Bigg\lbrace\left[(\bar{\beta} +\bar{\gamma}+2-\rho_{k-1})z +\rho_{k-1}-\bar{\beta}-1\right]u'_k(z)+\left[(\bar{\zeta}+\bar{\xi}+\bar{\alpha} \tau_k)z-\bar{\alpha} \tau_k-\bar{\xi}\right]u_k(z)\Bigg\rbrace=0.
\label{eq:53}
\end{equation}
Now, using the equations in \eqref{eq:50} to replace the terms \(u_k'\), \(z\,u_k'\), and subsequently making use of the freedom in choosing \( \rho_0\) to eliminate the term proportional to \(z\,u_k\), equation \eqref{eq:53} leads to:

\begin{equation}
\sum_{k=0}^N
C^{\pm}_k \left( R_k u_{k+1}(z) + Q_k u_k(z) + P_k u_{k-1}(z) \right) = 0,
\label{eq:54}
\end{equation}
with $\rho_0=n+\bar{\beta}-\tau_0-1$ and the terms $P_k,\,Q^{\pm}_k,R_k$ are given by 
\begin{equation}
R_k=\tau_{k+1} \tau_{k-n},\quad
Q_k^{\pm}=\tau_{k-n}(\tau_{k+1-n}-\bar{\beta}-\bar{\alpha})+\bar{\zeta},\quad
P_k=-\bar{\alpha}\,\tau_{k+1-n}.
\label{eq:55}
\end{equation}
However, since the polynomials \( u_k \) form a basis, the conditions \( P_0 = 0 \) and \( R_N = 0 \) must be satisfied simultaneously. The first condition allows us to determine the value of \( \tau_0 \), as it is fulfilled when
\begin{equation}
\tau_0=n-1,
\label{eq:56}  
\end{equation}
while the second condition is satisfied if
\begin{equation}
\tau_0=-N-1,\quad{\rm or}\quad
\tau_0=n-N.
\label{eq:57}  
\end{equation}
In this way, we observe that the value of \( N \) must be either 1 or \(-n\). Although both options lead to the same result, the calculation with \( N = -n \) is more complicated, as it requires summing over non-positive integers, and the number of terms to be summed is not fixed but depends on the order of the polynomial in question. For this reason, in the present work we will consider only the case \( N = 1 \), which leads to
\begin{equation}
\tau_0=n-1,\quad
\rho_0=\bar{\beta},\quad
R_k=(k-1)(k+n),\quad
Q_k^{\pm}=(k-1)(k-\bar{\beta}-\bar{\alpha})+\bar{\zeta},\quad
P_k=-\bar{\alpha}k.
\label{eq:58}  
\end{equation}
Once again, since $u_k$ form a basis, the following system is obtained:
\begin{equation}
\mathbf{M} \vec{C^{\pm}} = \vec{0}, \quad \text{with } 
\mathbf{M} = \begin{pmatrix}
R_0 & Q^{\pm}_1 \\
Q^{\pm}_0 & P_1
\end{pmatrix}, \quad
\vec{C^{\pm}} = \begin{pmatrix}
C^{\pm}_0 \\
C^{\pm}_1
\end{pmatrix}.
\label{eq:59}
\end{equation}
In order to avoid the trivial solution, the condition \(\det(\mathbf{M}) = R_0 P_1 - Q^{\pm}_1 Q^{\pm}_0 = 0\) must be satisfied. It is straightforward to verify that, using the values defined in \eqref{eq:40}, this condition holds. Therefore, the coefficients \( C_0^{\pm} \) and \( C_1^{\pm} \) are not independent and satisfy the following relation:
\begin{equation}
C^{\pm}_1Q^{\pm}_1+C^{\pm}_0 R_0=\bar{\zeta}^{\pm}C^{\pm}_1-nC^{\pm}_0=0.
\label{eq:60}
\end{equation}
It is important to emphasize that the above condition does not include the case \( n = 0 \), since for this value the condition \( 1 \leq \tau_k \) is not satisfied for \( k = 0, 1 \). However, this is a trivial case that admits the representation shown in equation~\eqref{eq:48}, provided that \( C_0^{\pm} = 0 \). In this way, we once again obtain two cases:
\begin{itemize}
\item[1] \textbf{Case $n=0$.} For this case, we have $C^{\pm}_0 = 0$. Then, the solution is given by
\begin{equation}
f^{\pm}(z) = C^{\pm}_1 \mbox{L}^{\bar{\beta}-1}_0(-\bar{\alpha}z),
\label{eq:61}    
\end{equation}
where $C^{\pm}_1$ will be determined by the normalization of the final solution.
\item[2] \textbf{Case $n\geq1$}. For this case, $C_0^{\pm}=\frac{\bar{\zeta}^{\pm}}{n}C_1^{\pm}$. Thus, $f^{\pm}$ is given by:
\begin{equation}
f^{\pm}(z) = C^{\pm}_1 \left(
\frac{\bar{\zeta}^{\pm}}{n}\mbox{L}^{\bar{\beta}}_{n-1}(-\bar{\alpha}z)+\mbox{L}^{\bar{\beta}-1}_{n}(-\bar{\alpha}z)\right).
\label{eq:62}
\end{equation}
\end{itemize}
It is important to emphasize that although \( \bar{\beta} \) is constant with respect to \( z \), its value depends on the energy and therefore on \( n \), being defined as:
\begin{equation}
\bar{\beta}_n\equiv\bar{\beta}(n)=  2\frac{\sqrt{k^2 v_y^2 - \left(E_n - k\,\nu\,v_{\rm t}\right)^2}}{\alpha v_x}.
\label{eq:63}
\end{equation}
Furthermore, since the argument of the associated Laguerre polynomials is given by \( -\bar{\alpha} z \), it is convenient to define the variable \( \theta \) as:
\begin{equation}
\theta(x)=-\bar{\alpha} z =\frac{2\mathcal{F}}{\alpha}{\rm e}^{-\alpha x}.
\label{eq:64}
\end{equation}
Thus, by substituting the solutions from Equations \eqref{eq:61} and \eqref{eq:62}, the eigenfunctions corresponding to the energy values given in Equation~\eqref{eq:46} are obtained as:
\begin{equation}
\Psi_n(x,y) = \frac{e^{ik_y y}}{\sqrt{2^{1- \delta_{n,0}}}} \begin{pmatrix} \psi^+_n(x) \\ \\ i\nu\mu\psi^-_n(x) \end{pmatrix},
\label{eq:65}
\end{equation}
with $\psi^{\pm}_n$ given by:
\begin{align}
\psi^+_n(x)&=\mathcal{N}_n {\rm e}^{-\frac{\theta}{2}}\theta^{\bar{\beta}_n/2}
\left(\frac{C_n\sqrt{\bar{\beta}_n+n}-\sqrt{n}}{\sqrt{\bar{\beta}_n+n}}\mbox{L}^{\bar{\beta}_n}_{n-1}\left( \theta_n \right)+\frac{\bar{\beta}_n}{\sqrt{n(\bar{\beta}_n+n)}}
(1-\delta_{n,0})\mbox{L}^{\bar{\beta}_n-1}_{n}
\left( \theta_n \right)\right),
\label{eq:66}\\
\psi^-_n(x)&=\mathcal{N}_n {\rm e}^{-\frac{\theta}{2}}\theta^{\bar{\beta}_n/2}
\left(\frac{\sqrt{\bar{\beta}_n+n}-\sqrt{n}C_n}{\sqrt{\bar{\beta}_n+n}}\mbox{L}^{\bar{\beta}_n}_{n-1}\left( \theta_n \right)+\frac{\bar{\beta}_n}{\sqrt{n(\bar{\beta}_n+n)}}
C_n(1-\delta_{n,0})\mbox{L}^{\bar{\beta}_n-1}_{n}
\left( \theta_n \right)\right),
\label{eq:67}
\end{align}
where the constants \( C_n \) and \( \mathcal{N}_n \) are defined as:
\begin{equation}
C_n=\frac{E_n-\nu v_{\rm t}k}{v_yk+\sqrt{v_y^2k^2-(E_n-\nu v_{\rm t}k)^2}},\quad
\mathcal{N}_n=\left(\frac{v_y k+\sqrt{v_y^2k^2-(E_n-\nu v_{\rm t}k)^2}}{2\left(v_yk+(E_n-\nu v_{rm t}k)\right)I_n}
\frac{\alpha\,n!}{\Gamma(\bar{\beta}_n+n)}
\right)^{\frac{1}{2}},
\label{eq:68}
\end{equation}
with the factor \( I_n \) denotes the following integral
\begin{equation}
I_n=2^{\delta_{n,0}}(1-\delta_{n,0})\frac{n!}{\Gamma(\bar{\beta}_n+n)}
\frac{1}{\sqrt{n(\bar{\beta}_n+n)}}
\int_0^{\infty}{\rm e}^{-\theta}\theta^{\bar{\beta}_n}
\mbox{L}_n^{\bar{\beta}_n-1}(\theta)\mbox{L}_{n-1}^{\bar{\beta}_n+1}(\theta)\,{\rm d}\theta.
\label{eq:69}
\end{equation}
We now present a brief discussion of these states and their corresponding energy levels.
\subsubsection{Discussion}
\begin{figure}[ht] 
\begin{center}
\includegraphics[width=0.55\textwidth]{Imagen1}
\subfigure[]{\includegraphics[width=8cm, height=5.7cm]{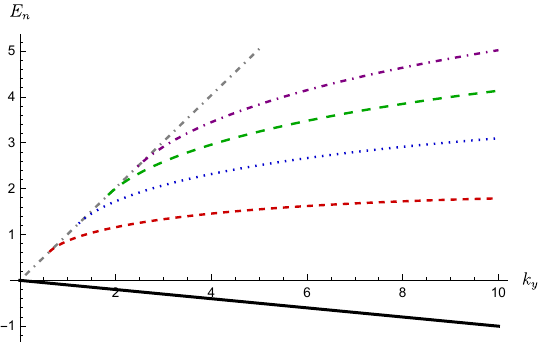}}
\subfigure[]{\includegraphics[width=8cm, height=5.7cm]{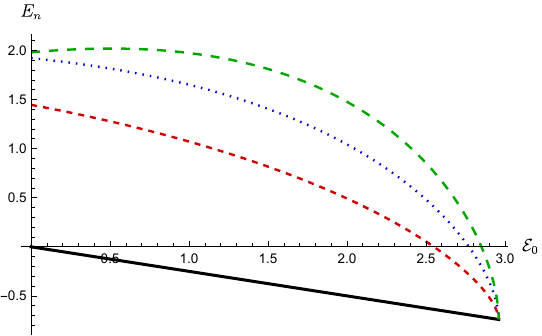}}
\subfigure[]{\includegraphics[width=8cm, height=5.7cm]{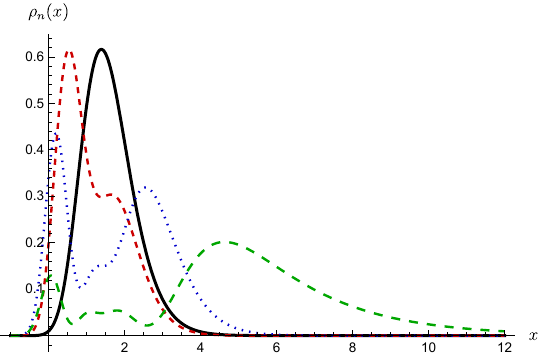}}
\subfigure[]{\includegraphics[width=8cm, height=5.7cm]{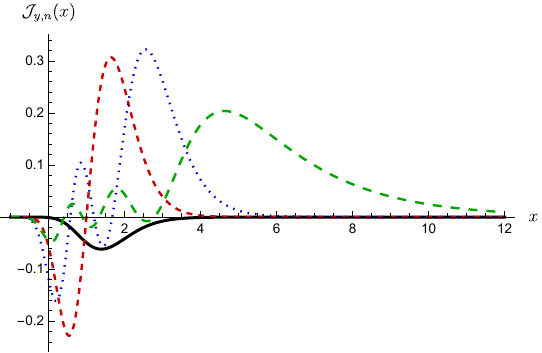}}
\caption{(a) Plots of the energy spectrum in \eqref{eq:46} as a function of the wavenumber $k_{y}$ and (b) as a function of the electric field strength $\mathcal{E}_{0}$. (c) Plot of the probability density $\rho_n(x)$ and (d) $y$-component current density $\mathcal{J}_{y,n}(x)$ corresponding to the eigenfunctions in \eqref{eq:65}. The values have been set as $k_{y}=2$, $\mathcal{B}_0=8$, $\alpha=\nu=\mu=1$, $\left\lbrace v_x,\;v_y,\;v_{\rm t},\;v_{\rm d}\right\rbrace=\left\lbrace 0.86,\;0.69,\;0.32,\;0.1\right\rbrace$.}
\label{F2}
\end{center}
\end{figure}

To begin, we must note that in this case it is necessary to impose a condition that ensures the energy levels in Equation \eqref{eq:48} are real and that the corresponding wave functions in Equation \eqref{eq:67} are square-integrable. This condition is given by
\begin{equation}
\kappa_n\equiv\kappa(n) >n\alpha,
\label{eq:70}
\end{equation}
where the dependence of $\kappa$ on $n$ arises from its relation to the energy (see Equation~\eqref{eq:12}).
This condition causes the energy spectrum to become discrete but finite, with the maximum number of excited states determined by the fulfilment of the previous condition. The spectrum exhibits a linear behavior in $k$ with a slope of $-v_d$ for the ground state, whereas it is nonlinear for the excited states. Moreover, since the square-integrability condition defines a domain in $k$, it can be observed that, as the number of excited states increases, this domain becomes narrower. Likewise, since the number of excited states is finite, it is possible to define an upper bound for all energy levels whose behavior is linear with respect to $k$, as indicated by the gray dashed line in Figure~\ref{F2}a. On the other hand, as in the previous case, when $\beta_\nu$ approaches $1$ (or $-1$) the energy levels tend to become degenerate at a value of $-v_d k$ (see Figure~\ref{F2}b). Regarding the energy current, it is also observed that the $y$-component is nonzero due to the tilt of the Dirac cones (see Figure \ref{F2}d).\\
\\
Finally, we would like to mention that, by using the identity
\[
\theta \, \mathrm{L}_{n-1}^{\beta+1}(\theta) = \beta \, \mathrm{L}_{n-1}^{\beta}(\theta) - n \, \mathrm{L}_n^{\beta-1}(\theta),
\]
our result coincides with that derived in \cite{Mojica-Zárate_2023, campa2024}, where a similarity transformation was employed to solve this problem.
\subsection{Hyperbolic fields}
For this example, we consider hyperbolic magnetic field profiles satisfying the relations $\mathbf{B} = \mathcal{B}_0 \, \mathrm{sech}^2(\alpha x)\, \hat{e}_z$ and $\mathbf{E} = \mathcal{E}_0 \, \mathrm{sech}^2(\alpha x)\, \hat{e}_x$, with $\mathcal{B}_0, \mathcal{E}_0, \alpha > 0$. Then, the $y$-component of the vector potential and the scalar potential can be chosen as $\mathcal{A}_y(x) = \frac{\mathcal{B}_0}{\alpha} \tanh(\alpha x)$ and $\phi(x) = -\frac{\mathcal{E}_0}{\alpha} \tanh(\alpha x)$, respectively. Additionally, let us define the following quantities:
\begin{equation}
\mathcal{F}=\sqrt{1-\beta_{\nu}^2}\frac{v_y}{v_x}\frac{\mathcal{B}_0}{\alpha},\quad\bar{a}=\frac{1}{2}\left(1+\frac{\sqrt{\epsilon}-\beta_\nu \kappa }{ \beta_\nu \mathcal{F}}\right),\quad
\bar{\delta} = \frac{\sqrt{(\mathcal{F}+\kappa)^2 - \epsilon}}{\alpha} + 1,\quad
\bar{\gamma}= \frac{\sqrt{(\mathcal{F}-\kappa)^2 - \epsilon}}{\alpha} + 1,
\nonumber
\end{equation}
\begin{equation}
\quad
\bar{\alpha} + \bar{\beta} = \frac{\sqrt{(\mathcal{F} + \kappa)^2 - \epsilon}}{\alpha} + \frac{\sqrt{(\mathcal{F} - \kappa)^2 - \epsilon}}{\alpha},\quad
\bar{\alpha} \, \bar{\beta} = \frac{-\mathcal{F}^2 - \epsilon + \kappa^2 + \sqrt{-\epsilon + (\mathcal{F} - \kappa)^2} \, \sqrt{-\epsilon + (\mathcal{F} + \kappa)^2}}{2 \alpha^2},
\nonumber
\end{equation}
\begin{equation}
\bar{q}^{\pm} = \frac{1}{2} \left( 
\bar{a} \left( \bar{\alpha} + \bar{\beta} \right) 
+ 2\bar{a} \bar{\alpha} \bar{\beta} 
- (\bar{\gamma} - 1) 
\pm \frac{\sqrt{1 - \beta_\nu^2} \sqrt{\epsilon} }{ \alpha \beta_\nu } 
\right).
\label{eq:72}
\end{equation}
Thus, by using the change of variable
\begin{equation}
z = \frac{1 + \tanh(\alpha x)}{2},
\label{eq:73}
\end{equation}
the system of differential equations in \eqref{eq:10} becomes:
\begin{equation}
\left[\frac{{\rm d}^2}{{\rm d}z^2}+\left(\frac{z^2 - 2\bar{a} z+ \bar{a}}{z (z - 1) (z - \bar{a})}\right)\frac{{\rm d}}{{\rm d}z}+\left(\frac{A_0 + A_1 z + A_2 z^2 + A_3 z^3}{z^2 (z - 1)^2 (z - \bar{a})}\right)\right]\phi^{\pm}(z)=0,
\label{eq:74}
\end{equation}
where we take $\psi^{\pm}(x)=\phi^{\pm}(z)$ and the coefficients $A_i$ are given by
\begin{equation}
A_0 = \frac{1}{4} \bar{a} (\bar{\gamma} - 1)^2, \quad
A_1 = -\frac{1}{4} \bar{\gamma} \left(2 \bar{a} (\bar{\alpha} + \bar{\beta}) + \bar{\gamma} - 4\right) + \bar{q}^{\pm} - \frac{3}{4}, \nonumber
\end{equation}
\begin{equation}
A_2 = \frac{1}{4} \left(\bar{a} (\bar{\alpha} + \bar{\beta})(\bar{\alpha} + \bar{\beta} + 2) + 2 \bar{\gamma} (\bar{\alpha} + \bar{\beta}) - 4 \bar{\alpha} \bar{\beta} - 2 \bar{\alpha} - 2 \bar{\beta} - 2 \bar{\gamma} - 4 \bar{q}^{\pm} + 2\right), \quad
A_3 = -\frac{1}{4} (\bar{\alpha} - \bar{\beta})^2.
\label{eq:75}
\end{equation}
Note that $z \to 0$ as $x \to -\infty$. Furthermore, in this limit, the differential operator in \eqref{eq:74} takes the form \mbox{$\frac{{\rm d^2}}{{\rm d}z^2}
+\frac{1}{z}\frac{{\rm d}}{{\rm d}z}-\frac{(\bar{\gamma}-1)^2}{4z^2}$}, so asymptotically the solution must behave like $\phi^{\pm}(z)\propto z^{\frac{\gamma -1}{2}}$ in this region. On the other hand, when $x \to \infty$, we have $z \to 1$, and in this domain limit, the differential operator becomes $\frac{{\rm d^2}}{{\rm d}z^2}
+\frac{1}{z-1}\frac{{\rm d}}{{\rm d}z}-\frac{(\bar{\delta}-1)^2}{4(z-1)^2}$. Therefore, in this region, the solution should asymptotically behave like $\phi^{\pm}(z)\propto (1-z)^{\frac{\delta -1}{2}}$. Based on all the above, we can propose that the solution to equation \eqref{eq:74} takes the form
\begin{equation}
\phi^{\pm}(z)=z^{\frac{\gamma -1}{2}} (z-1)^{\frac{\delta -1}{2}}
f^{\pm}(z).
\label{eq:76}
\end{equation}
Substituting this ansatz in \eqref{eq:74}, a new differential equation for $f^{\pm}$ is obtained:
\begin{equation}
\left[\frac{{\rm d}^2}{{\rm d}z^2} 
+\frac{(\bar{\alpha} + \bar{\beta} + 1)z^2+\left( \bar{a} (-\bar{\gamma} - \bar{\delta}) - \bar{\alpha} - \bar{\beta} + \bar{\delta} - 1 \right)z+\bar{a}\bar{\gamma}}{z (z - 1) (z - \bar{a})}\frac{{\rm d}}{{\rm d}z}
+\frac{\bar{\alpha} \bar{\beta} z - \bar{q}^{\pm}}{z (z - 1) (z - \bar{a})}\right]\phi^{\pm}(z)=0.
\label{eq:77}
\end{equation}
The above equation is known as the general form of the Heun differential equation \cite{MELIKDZHANIAN2021125037}. Consequently, the function $f^{\pm}(z) = \text{HeunG}(\bar{\alpha},\bar{\beta},\bar{\gamma},\bar{\delta},\bar{q}^{\pm};z)$. As in previous cases, this solution $\psi^{\pm}$ will be square integrable if $f^{\pm}$ is a polynomial of finite degree $n$. For this to hold, the following condition must be satisfied:
\begin{equation}
n^2 + (\bar{\alpha} + \bar{\beta}) n + \bar{\alpha} \bar{\beta} = 0.
\label{eq:78}
\end{equation}
Note that this equation has solutions $\bar{\alpha} = -n$ or $\bar{\beta} = -n$. However, given the definitions of these parameters in \eqref{eq:72}, both conditions cannot occur simultaneously. Nevertheless, since these parameters can be freely interchanged without affecting any of the previous equations, we may assume, without loss of generality, that $\bar{\alpha}=-n$ and $\bar{\beta}\neq-n$ (or vice versa). Moreover, as in the previous cases, this condition acts as a quantization relation. In contrast to the quantization conditions for constant and exponential fields, this condition is not linear in $n$, but quadratic. As a result, there are four independent solutions for the energy, two of which are:
\begin{align}
E_n&=\nu k v_{\rm t}- \frac{v_y\beta_{\nu} k \mathcal{F}^2}
{(\mathcal{F} - n \alpha )^2 + \beta_{\nu}^2 (2\mathcal{F} - n \alpha) n\alpha}\nonumber\\
&\quad+\mu 
\frac{(1-\beta_{\nu}^2)(\mathcal{F}-n\alpha )^2
}{(\mathcal{F} - n\alpha)^2+
\beta_{\nu}^2 (2\mathcal{F}-n\alpha)n\alpha }\sqrt{
\frac{
n\alpha(2\mathcal{F}- n\alpha)
\left(
v_x^2 \left((\mathcal{F} - \alpha n)^2+\beta_{\nu}^2 (2\mathcal{F} - \alpha n)n\alpha\right)-(1-\beta_{\nu}^2)v_y^2k^2
\right)
}{(1 - \beta_{\nu}^2)(\mathcal{F} - \alpha n)^2}},
\label{eq:79}
\end{align}
with the band index $\mu=\pm 1$ and $n\in\mathbb{Z}_{\geq0}$. For these solutions to be well-defined and for the spectrum to remain real, it is necessary to impose the following condition:
\begin{equation}
n\alpha<\mathcal{F}.
\label{eq:80}
\end{equation}
Recalling that $\mathcal{F}$ is positive definite, for a given value of $\mathcal{F}$, this condition imposes an upper bound on the number of excited states. Now, returning to the other two possible energy spectra obtained by solving Equation \eqref{eq:78}, it can be observed that they result from implementing the transformation $n\alpha \rightarrow -n\alpha$ in Equation \eqref{eq:79}. However, this also modifies the condition required for the energy spectrum to be well-defined and real, demanding now that $\mathcal{F} < (v_y / v_x) k$. To ensure that this condition is always satisfied, the fields would need to vanish or $k \rightarrow \infty$, which contradicts the goal of obtaining stationary states. For this reason, we will henceforth consider only the energy spectrum given in Equation \eqref{eq:79}, along with the condition specified in Equation \eqref{eq:80}. In this way, it is easy to verify that $\bar{\alpha}$ and $\bar{\beta}$ are given by
\begin{equation}
\bar{\alpha}=-n,\quad
\bar{\beta}=\bar{\alpha}+\frac{2\mathcal{F}}{\alpha}.
\label{eq:81}
\end{equation}
On the other hand, similarly to the previous cases, the polynomial solution for $f^{\pm}$ can be written in terms of Jacobi polynomials, as we will show below.
\subsubsection{Polynomial solutions of the general Heun differential equation}
By considering that the polynomial solution to the general Heun differential equation can be expressed in a basis of Jacobi polynomials, the following condition must be satisfied:
\begin{equation}
f^{\pm}(z) = \sum_{k=0}^n A^{\pm}_k z^k = \sum_{k=0}^N C^{\pm}_k\, u_k(z), \quad
u_k(z) = \mathrm{P}_{\tau_0 + k}^{(a_0,b_0)}\left( s_0(z + z_0) \right),
\label{eq:82}
\end{equation}
where $\tau_0$, $a_0$, $b_0$, $s_0$ and $z_0$ are constants that can be freely chosen, provided they ensure the convergence of the solution. Moreover, the functions $u_k$ satisfy the following differential equation:
\begin{equation}
\frac{ \left(1 - s_0^2 (z + z_0)^2 \right)}{s_0^2} u_k''(z)
+ \frac{b_0 - a_0 - s_0 (a_0 + b_0 + 2)(z + z_0)}{s_0} u_k'(z) 
+ \tau_k (a_0 + b_0 + \tau_{k+1}) u_k(z) 
= 0,
\label{eq:83}
\end{equation}
as well as the recurrence relations given by
\begin{align}
u_k'(z) 
&=\frac{s_0\left(\tau_k \left( a_0 - b_0 - s_0 (z + z_0)(a_0 + b_0 + 2\tau_k) \right) u_k(z) 
+ 2 (a_0 + \tau_k)(b_0 + \tau_k) u_{k-1}(z)\right)}{ \left(1 - s_0^2 (z + z_0)^2 \right)(a_0 + b_0 + 2\tau_k)},\nonumber\\
z\, u_k(z) &=\frac{1}{2} \left(
\frac{2\, (a_0 + \tau_k)(b_0 + \tau_k)}{(a_0 + b_0 + 2\tau_k)(1 + a_0 + b_0 + 2\tau_k)}\, u_{k-1}(z) 
+ \frac{2 (1 + b_0)(a_0 + b_0) + 4 \tau_k (1 + a_0 + b_0 + \tau_k)}{(a_0 + b_0 + 2\tau_k)(2 + a_0 + b_0 + 2\tau_k)}\, u_k(z)\right.\nonumber\\
&\quad\left.+ \frac{2\, (1 + \tau_k)(1 + a_0 + b_0 + \tau_k)}{(1 + a_0 + b_0 + 2\tau_k)(2 + a_0 + b_0 + 2\tau_k)}\, u_{k+1}(z)
\right).
\label{eq:84}
\end{align}
Thus, by inserting the ansatz from Equation \eqref{eq:82} into the general Heun differential equation \eqref{eq:77}, and replacing the second derivative of $u_k$ by means of the Equation \eqref{eq:83}, one obtains, after some algebraic manipulation, the following expression:
\begin{equation}
\sum_{k=0}^N C^{\pm}_k\,\left[\left(A_0+A_1z+A_2z^2+A_3z^3+A_4z^4\right)u'_k(z)+\left(D_0+D_1z+D_2z^2+D_3z^3\right)u_k(z)\right]=0,
\label{eq:85}
\end{equation}
where the constants $A_i$ and $D_i$ are given by
\begin{align}
A_0&=- \bar{a} (1 - s_0^2 z_0^2)\, \bar{\gamma}
,\nonumber\\
A_1&=(1 - s_0^2 z_0^2)(\bar{\gamma} - 1) 
+ \bar{a} \left[
2 + \bar{\alpha} + \bar{\beta} 
- s_0 \left(
a_0 - b_0 
+ s_0 z_0 \left(
2 + a_0 + \left( b_0 + z_0 (2 + \bar{\alpha} + \bar{\beta}) - 2 \bar{\gamma} \right)
\right)
\right)
\right]
,\nonumber\\
A_2&=-1 - \bar{\alpha} - \bar{\beta} 
+ s_0 \left[
(1 + \bar{a})\, b_0 (s_0 z_0 - 1) 
+ a_0 \left((1 + \bar{a})(1 + s_0 z_0) - \bar{a} s_0\right) 
+ s_0 z_0 \left(4 + z_0 (1 + \bar{\alpha} + \bar{\beta}) - 2 \bar{\gamma}\right)\right.\nonumber\\ 
&\left.+ \bar{a} s_0 \left(\bar{\gamma} - 2 - 2 z_0 (1 + \bar{\alpha} + \bar{\beta})\right)
\right]
,\nonumber\\
A_3&=s_0 \left(
b_0 + s_0 + 2 \bar{a} s_0 
+ a_0 \left(s_0 (1 + \bar{a} - z_0)-1 \right) 
+ s_0 \left[
2 + b_0 (1 + \bar{a} - z_0) 
+ (2 z_0- \bar{a}) (\bar{\alpha} + \bar{\beta}) 
-2\bar{a} 
- \bar{\gamma}
\right]
\right)
,\nonumber\\
A_4&=s_0^2 \left(\bar{\alpha} + \bar{\beta} - a_0 - b_0 - 1\right)
,\nonumber\\
D_0&=\bar{q}^{\pm} \left(1 - s_0^2 z_0^2\right),\nonumber\\
D_1&=-\bar{\alpha} \bar{\beta} + 
s_0^2 \left(z_0 \left(-2 \bar{q}^{\pm} + z_0 \bar{\alpha} \bar{\beta}\right) + 
\bar{a} (k + \tau_0) (1 + a_0 + b_0 + k + \tau_0)\right)
,\nonumber\\
D_2&=s_0^2 \left(- (1 + \bar{a}) k^2 - \bar{q}^{\pm} + 
2 z_0 \bar{\alpha} \bar{\beta} - (1 + \bar{a}) \tau_0 (1 + a_0 + b_0 + \tau_0) - (1 + \bar{a}) k (1 + a_0 + b_0 + 2 \tau_0) \right)
,\nonumber\\
D_3&=s_0^2 \left(k^2 + \bar{\alpha} \bar{\beta} + \tau_0 (1 + a_0 + b_0 + \tau_0) + k (1 + a_0 + b_0 + 2 \tau_0) \right).
\label{eq:86}
\end{align}
Using the freedom in the choice of $s_0$ and $z_0$ to eliminate the terms proportional to $u'_k$ and $z u'_k$ in Equation \eqref{eq:85}, and the freedom in $a_0$ and $b_0$ to ensure that the remaining terms share a common factor with $u_k$, we propose:
\begin{equation}
z_0 =-\frac{1}{2}, \quad
s_0 = -\frac{1}{z_0}=2, \quad
a_0 = \bar{\delta} - 1, \quad
b_0 = \bar{\gamma} - 1.
\label{eq:87}
\end{equation}
It is worth mentioning that the previous conditions are not the only ones that lead to the required constraints. In fact, an equally valid set is given by $z_0=-\frac{1}{2}$, $s_0=\frac{1}{z_0}=-2$, $a_0=\bar{\gamma}-1$, $b_0 = \bar{\delta} - 1$. However, with these values, the final result will differ by a global phase, which is physically irrelevant. For this reason, and without loss of generality, we will only consider the case described by the conditions in Equation \eqref{eq:87}. Thus, Equation \eqref{eq:85}, after some algebraic manipulation, leads to
\begin{align}
\sum^N_{k=0}C^{\pm}_k&\left[
-4z(z-1)\, (1 + \bar{\alpha} + \bar{\beta} - \bar{\gamma} - \bar{\delta})\, u'_k(z)
- 4 \left(
\bar{q}^{\pm} + \bar{a} (k + \tau_0)(-1 + k + \bar{\gamma} + \bar{\delta} + \tau_0)\right.\right.\nonumber\\
&\left.\left.- z \left(k^2 + \bar{\alpha} \bar{\beta} + \tau_0(-1 + \bar{\gamma} + \bar{\delta} + \tau_0)
+ k(-1 + \bar{\gamma} + \bar{\delta} + 2\tau_0)
\right)\right)u_k(z)
\right]=0.
\label{eq:88}
\end{align}
Nevertheless, using the recurrence relations in \eqref{eq:84} to replace the first derivative and the terms involving $z u_k(z)$, the previous equation transforms into
\begin{equation}
\sum_{k=0}^N
C^{\pm}_k \left( R_k u_{k+1}(z) + Q^{\pm}_k u_k(z) + P_k u_{k-1}(z) \right) = 0,
\label{eq:89}
\end{equation}
where $R_k$, $P_k$, and $Q^{\pm}_k$ are given by
\begin{align}
R_k&=\frac{4 (1 + k + \tau_0) (k + \bar{\alpha} + \tau_0) (k + \bar{\beta} + \tau_0) (1 + k + \bar{\alpha} + \bar{\beta} + \tau_0)}{(1 + 2 k + \bar{\alpha} + \bar{\beta} + 2 \tau_0) (2 + 2 k + \bar{\alpha} + \bar{\beta} + 2 \tau_0)}
,\nonumber\\
Q^{\pm}_k&= \frac{1}{4} \Bigg(
-2 \big(-4 + (-4 + 8 \bar{a}) k^2 + 8 \bar{q}^{\pm} + (\bar{\alpha} - \bar{\beta})^2 
+ 4 (-1 + 2 \bar{a}) k (1 + \bar{\alpha} + \bar{\beta}) \big)\nonumber\\
&\quad+ 4 (-2 + \bar{\alpha} + \bar{\beta}) \bar{\gamma} 
- 8 (-1 + 2 \bar{a}) (1 + 2 k + \bar{\alpha} + \bar{\beta}) \tau_0 + 8 (1 - 2 \bar{a}) \tau_0^2\nonumber\\
&\quad+ \frac{(-2 + \bar{\alpha} - \bar{\beta})(2 + \bar{\alpha} - \bar{\beta})(\bar{\alpha} + \bar{\beta})(2 + \bar{\alpha} + \bar{\beta} - 2 \bar{\gamma})}{2 k + \bar{\alpha} + \bar{\beta} + 2 \tau_0}\nonumber\\
&\quad- \frac{(-2 + \bar{\alpha} - \bar{\beta})(2 + \bar{\alpha} - \bar{\beta})(\bar{\alpha} + \bar{\beta})(2 + \bar{\alpha} + \bar{\beta} - 2 \bar{\gamma})}{2 + 2 k + \bar{\alpha} + \bar{\beta} + 2 \tau_0}
\Bigg),\nonumber\\
P_k&=\frac{4 (1 + k + \bar{\alpha} + \tau_0) (1 + k + \bar{\beta} + \tau_0) (1 + k + \bar{\alpha} + \bar{\beta} - \bar{\gamma} + \tau_0) (-1 + k + \bar{\gamma} + \tau_0)}{(2 k + \bar{\alpha} + \bar{\beta} + 2 \tau_0) (1 + 2 k + \bar{\alpha} + \bar{\beta} + 2 \tau_0)}.
\label{eq:90}
\end{align}
Now, being that the functions $u_k(z)$ form a basis, it is necessary to demand that $P_0=0$ and $R_N=0$. Because the numerator of $P_0$ is a fourth-order polynomial in $\tau_0$, there will be several solutions; however, only one of them yields an integer value, corresponding to $\tau_0=-\bar{\alpha}-1 =n-1$. Similarly, the condition $R_N=0$ yields four different solutions for $\tau_0$, but only two of them are integers, corresponding to
\begin{equation}
\tau_0=n-N,\quad{\rm or}\quad 
\tau_0=-N-1.
\label{eq:91}
\end{equation}
Thus, $N$ can only take the values $1$ and $-n$. However, as in the previous cases, the value $-n$ leads to the same result but involves more complicated calculations. For this reason, we will only consider the value $N=1$. In this way, we obtain that
\begin{align}
R_k&=\frac{4 (k - 1) (k - \bar{\alpha}) (k + \bar{\beta}) (k - 1 - \bar{\alpha} + \bar{\beta})}{(2 k - 1 - \bar{\alpha} + \bar{\beta}) (2 k - \bar{\alpha} + \bar{\beta})}
,\nonumber\\
Q^{\pm}_k&= -\frac{1}{\left(-2 + 2k - \bar{\alpha} + \bar{\beta}\right)\left(2k - \bar{\alpha} + \bar{\beta}\right)} \times \nonumber\\
&\quad 4\Bigg[
(-2 + 4\bar{a})k^4 
- 4(-1 + 2\bar{a})k^3(1 + \bar{\alpha} - \bar{\beta})  \nonumber \\
&\quad + (2 + \bar{\alpha} - \bar{\beta})\left( 
\bar{q}^{\pm}(\bar{\alpha} - \bar{\beta}) 
+ (1 + \bar{\alpha})\bar{\beta}(1 - \bar{a}\bar{\alpha} + \bar{a}\bar{\beta}) 
- \bar{\beta}\bar{\gamma}
\right) \nonumber \\
&\quad + k^2\left(
-4 + 4\bar{q}^{\pm} - 5\bar{\alpha} - 2\bar{\alpha}^2 + 7\bar{\beta} 
+ 6\bar{\alpha}\bar{\beta} - 2\bar{\beta}^2 \right. \nonumber \\
&\quad + \bar{a}(4 + 10\bar{\alpha} + 5\bar{\alpha}^2 
- 14(1 + \bar{\alpha})\bar{\beta} + 5\bar{\beta}^2) 
- (-2 + \bar{\alpha} + \bar{\beta})\bar{\gamma}
\bigg) \nonumber \\
&\quad - k(1 + \bar{\alpha} - \bar{\beta})\bigg(
4\bar{q}^{\pm} + \bar{a}\bar{\alpha}^2 
+ \bar{\beta}(3 + \bar{a}(-6 + \bar{\beta}) - \bar{\gamma}) \nonumber \\
&\quad + 2(-1 + \bar{\gamma}) 
- \bar{\alpha}(1 - 2\bar{\beta} + \bar{a}(-2 + 6\bar{\beta}) + \bar{\gamma})
\bigg)
\Bigg],\nonumber\\
P_k&=\frac{4 k (k - \bar{\alpha} + \bar{\beta}) (k + \bar{\beta} - \bar{\gamma}) (k - 2 - \bar{\alpha} + \bar{\gamma})}{
(2k-2 - \bar{\alpha} + \bar{\beta})(2k-1 - \bar{\alpha} + \bar{\beta})}.
\label{eq:92}
\end{align}
As the functions $u_k$ form a basis, the following system is obtained:
\begin{equation}
\mathbf{M} \vec{C^{\pm}} = \vec{0}, \quad \text{with } 
\mathbf{M} = \begin{pmatrix}
R_0 & Q^{\pm}_1 \\
Q^{\pm}_0 & P_1
\end{pmatrix}, \quad
\vec{C^{\pm}} = \begin{pmatrix}
C^{\pm}_0 \\
C^{\pm}_1
\end{pmatrix}.
\label{eq:93}
\end{equation}
Once again, to avoid the trivial solution, it is necessary that \(\det(\mathbf{M}) = R_0 P_1 - Q^{\pm}_1 Q^{\pm}_0 = 0\), which can be easily verified by a direct substitution of the parameters involved. In this way, $C^{\pm}_1$ and $C^{\pm}_0$ are not linearly independent and satisfy the following relation:
\begin{equation}
C^{\pm}_1Q^{\pm}_1+C^{\pm}_0 R_0=0.
\label{eq:94}
\end{equation}
Equations \eqref{eq:83} and \eqref{eq:94} lead to the following two cases:
\begin{itemize}
\item[1] \textbf{Case $n=0$}. In this case, the solution is given by a polynomial of degree zero, and therefore $C^{\pm}_0=0$. In this way, the functions $f^{\pm}$ are given by
\begin{equation}
f^{\pm}(z) = C^{\pm}_1 \, \mathrm{P}^{(\bar{\delta}-1,\bar{\gamma}-1)}_0(2z-1),
\label{eq:95}
\end{equation}
where $C_1^{\pm}$ is determined by the normalization procedure.
\item[2] \textbf{Case $n\geq1$}. For this case, $R_0\neq0$, and thus we must have $C_0^{\pm}=\frac{\mathcal{F} \sqrt{\varepsilon} \left(1 \pm \sqrt{1 - \beta_{\nu}^2} \right)}
{n \alpha \left(n \alpha-2 \mathcal{F}\right) \beta_{\nu}}
C_1^{\pm}$. Then, $f^{\pm}$ takes the form
\begin{equation}
f^{\pm}(z)=C^{\pm}_1\left(
\frac{\mathcal{F} \sqrt{\varepsilon} \left(1 \pm \sqrt{1 - \beta_{\nu}^2} \right)}
{n \alpha \left(n \alpha-2 \mathcal{F}\right) \beta_{\nu}}
\mathrm{P}^{(\bar{\delta}-1,\bar{\gamma}-1)}_{n-1}
+\mathrm{P}^{(\bar{\delta}-1,\bar{\gamma}-1)}_{n}\right),
\label{eq:96}
\end{equation}
where $C_1^{\pm}$ is fixed by the final normalization of the function.
\end{itemize}
It is important to note that $\epsilon$, $\bar{\gamma}$ and $\bar{\delta}$ depend on the energy and therefore on $n$, i.e., $\epsilon=\epsilon_n$, $\bar{\gamma} = \bar{\gamma}_n$ and $\bar{\delta} = \bar{\delta}_n$. Moreover, if we define $\theta$ as the argument of the Laguerre polynomials, we have
\begin{equation}
\theta\equiv\theta(x)=2z-1={\rm tanh}(\alpha x).
\label{eq:97}
\end{equation}
Therefore, the corresponding eigenfunctions for the energies in Equation \eqref{eq:79} are given by
\begin{equation}
\Psi_n(x,y) = \frac{e^{ik_y y}}{\sqrt{2^{1- \delta_{n,0}}}} \begin{pmatrix} \psi^+_n(x) \\ \\ i\nu\mu\psi^-_n(x) \end{pmatrix},
\label{eq:98}
\end{equation}
where $\psi^{\pm}_n$ turn out to be:
\begin{align}
\psi^+_n(x)&=\mathcal{N}_n (1-\theta)^{\frac{\bar{\delta}_n-1}{2}}(1+\theta)^{\frac{\bar{\gamma}_n-1}{2}}
\left(w^{(1)}_n\mbox{P}^{(\bar{\delta}_n-1,\bar{\gamma}_n-1)}_{n-1}
\left(\theta \right)+w^{(2)}_n\mbox{P}^{(\bar{\delta}_n-1,\bar{\gamma}_n-1)}_{n}
\left(\theta \right)\right),
\label{eq:99}\\
\psi^-_n(x)&=\mathcal{N}_n (1-\theta)^{\frac{\bar{\delta}_n-1}{2}}(1+\theta)^{\frac{\bar{\gamma}_n-1}{2}}
\left(w^{(3)}_n\mbox{P}^{(\bar{\delta}_n-1,\bar{\gamma}_n-1)}_{n-1}
\left(\theta \right)+w^{(4)}_n\mbox{P}^{(\bar{\delta}_n-1,\bar{\gamma}_n-1)}_{n}
\left(\theta \right)\right),
\label{eq:100}
\end{align}
with the functions \( w^{(i)}_n \) given by  
\begin{equation}
w_n^{(1)}=(1-\delta_{n,0})\left(
\frac{(\bar{\delta}_n+n-1)(\bar{\gamma}_n+n-1)}{n(\bar{\delta}_n+\bar{\gamma}_n+n-2)}\right)^{\frac{1}{2}}M_n,\quad
w_n^{(2)}=-\frac{\mu\beta_{\nu}}{1+\sqrt{1-\beta_{\nu}^2}}M_n,\nonumber
\end{equation}
\begin{equation}
w_n^{(3)}=-\frac{\mu\beta_{\nu}}{1+\sqrt{1-\beta_{\nu}^2}}(1-\delta_{n,0})\left(
\frac{(\bar{\delta}_n+n-1)(\bar{\gamma}_n+n-1)}{{n(\bar{\delta}_n+\bar{\gamma}_n+n-2)}}\right)^{\frac{1}{2}}M_n,\quad
w_n^{(4)}=M_n,
\label{eq:101}
\end{equation}
being $M_n$ written as
\begin{equation}
M_n=2^{-\frac{\bar{\delta}_n+\bar{\gamma}_n-2}{2}}
\left(\frac{2\alpha(\bar{\delta}_n-1)(\bar{\gamma}_n-1)n!
\Gamma(\bar{\delta}_n+\bar{\gamma}_n+n-1)}
{(\bar{\delta}_n+\bar{\gamma}_n-2)\Gamma(\bar{\delta}_n+n)\Gamma(\bar{\gamma}_n+n)}\right)^{\frac{1}{2}},
\label{eq:102}
\end{equation}
and \( \mathcal{N}_n \) denotes the normalization constant, explicitly given by  
\begin{equation}
\mathcal{N}_n=\left( \frac{1+\sqrt{1+\beta_{\nu}^2}}{2(1-2^{\delta_{n,0}}\mu\beta_{\nu}(1-\delta_{n,0})I_n)},\right)^{\frac{1}{2}},
\label{eq:103}
\end{equation}
with \( I_n \) being the integral defined by  
\begin{equation}
I_n=\frac{M_n^2}{\alpha}\left(
\frac{(\bar{\delta}_n+n-1)(\bar{\gamma}_n+n-1)}{{n(\bar{\delta}_n+\bar{\gamma}_n+n-2)}}\right)^{\frac{1}{2}}\int^1_{-1} \, (1-\theta)^{\bar{\delta}_n-2}(1+\theta)^{\bar{\gamma}_n-2}
{\rm P}_n^{(\bar{\delta}_n-1,\bar{\gamma}_n-1)}(\theta)
{\rm P}_{n-1}^{(\bar{\delta}_n-1,\bar{\gamma}_n-1)}(\theta)\,{\rm d}\theta.
\label{eq:104}
\end{equation}
\subsubsection{Discussion}
\begin{figure}[ht] 
\begin{center}
\includegraphics[width=0.55\textwidth]{Imagen1}
\subfigure[]{\includegraphics[width=8cm, height=5.7cm]{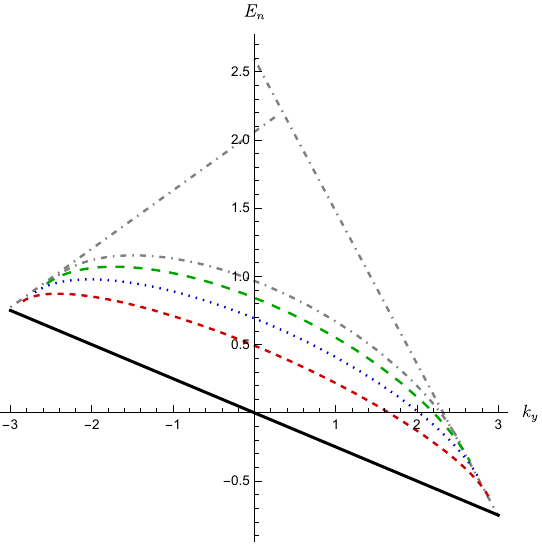}}
\subfigure[]{\includegraphics[width=8cm, height=5.7cm]{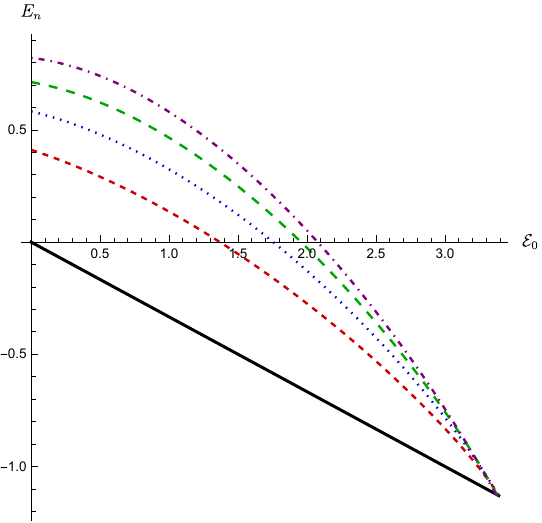}}
\subfigure[]{\includegraphics[width=8cm, height=5.7cm]{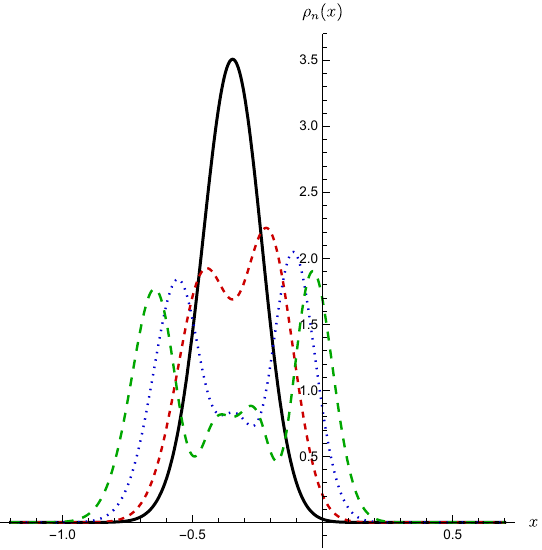}}
\subfigure[]{\includegraphics[width=8cm, height=5.7cm]{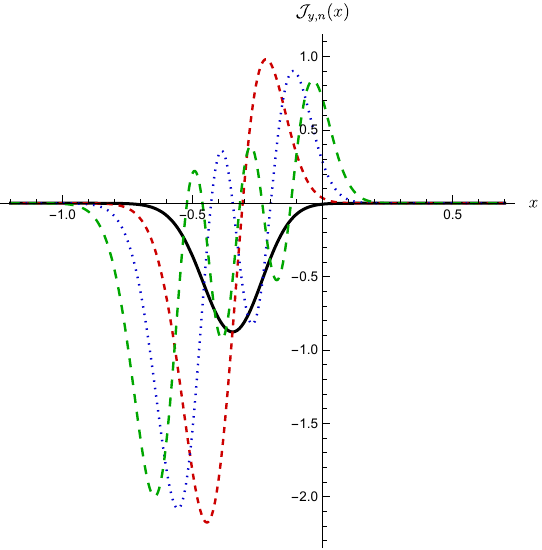}}
\caption{(a) Plots of the energy spectrum in \eqref{eq:79} as a function of the wavenumber $k_{y}$ and (b) as a function of the electric field strength $\mathcal{E}_{0}$. (c) Plot of the probability density $\rho_n(x)$ and (d) $y$-component current density $\mathcal{J}_{y,n}(x)$ corresponding to the eigenfunctions in \eqref{eq:98}. The values have been set as $k_{y}=1$, $\mathcal{B}_0=3$, $\alpha=\nu=\mu=1$, $\left\lbrace v_x,\;v_y,\;v_{\rm t},\;v_{\rm d}\right\rbrace=\left\lbrace 0.0524,\;0.785,\;-0.345,\;0.35\right\rbrace$.}
\label{F3}
\end{center}
\end{figure}

For this case, it is necessary to impose an additional condition to guarantee the square-integrability of the states described in Equation \eqref{eq:98}, which is given by
\begin{equation}
|\kappa_n|<\frac{(\mathcal{F}-n\alpha)^2}{\mathcal{F}}.
\label{eq:106}
\end{equation}
Thus, in addition to having a finite number of states due to the condition in Equation \eqref{eq:80}, we obtain a well-defined domain in $k$ for each energy level. Consequently, the resulting spectrum is discrete, finite, dispersive, and bounded. However, although the bounds remain linear in $k$, they do not exhibit the same slope on both sides. Furthermore, the domain tends to decrease as the number of excited states increases (see Figure~\ref{F3}a).\\
\\
On the other hand, when the electric field approaches its critical value, the right-hand side of the inequality in Equation \eqref{eq:106} is affected, thereby modifying the regions where the existence of excited states is allowed. Thus, when plotting the energy as a function of $\mathcal{E}_0$, the states tend to become degenerate again, approaching the value $-k v_{\mathrm{d}}$ (see Figures~\ref{3}b). Finally, as in previous cases, both the probability densities and probability currents depend on the band and valley indices, with the latter showing a nonzero value in the ground state (see Figures~\ref{3}c and \ref{3}d).
\section{Conclusions}
\label{4}
In this work, we analyzed the dynamics of Dirac materials subjected to position-dependent electric and magnetic fields by decoupling the corresponding system of differential equations. The resulting equations, although not adopting the standard eigenvalue form, exhibit deep connections with Heun equations. The quantization conditions derived from finite polynomial solutions of these equations allowed us to obtain discrete energy spectra for each field configuration considered.

Across the different scenarios studied—from constant and exponential profiles to hyperbolic field configurations—we observed a consistent set of behaviors and phenomena. In all cases, the energy spectrum features a term with linear dependence on the wave number \(k\) in the \(y\)-direction. For the ground state, the slope of this linear term is universally given by \(-v_{\mathrm{d}}\), corresponding to the ratio between the amplitudes of the electric and magnetic fields. For the excited states, the behavior varies depending on the specific field configuration. In the first two cases, the slope of the linear component remains constant across all energy levels. However, in hyperbolic configurations, the slope depends on the quantum number \(n\), and only the ground state retains a slope equal to \(-v_{\mathrm{d}}\). Therefore, we can conclude that the presence of the electric field generates dispersive energy levels, characterized by a linear dependence on \( k \).

The presence of a critical electric field strength, defined by \(\mathcal{E}_0 = \mathcal{B}_0 (v_y - \nu v_{\rm t})\), plays a crucial role. As the electric field approaches this critical value, all energy levels collapse to the unique value \(E = -k v_{\mathrm{d}}\), indicating a degeneracy point consistently observed in each configuration (see Figures~\ref{F1}b, \ref{F2}b, and \ref{F3}b). In the exponential and hyperbolic cases, it is necessary to impose additional conditions to ensure the square-integrability of the wavefunctions. These constraints limit the number of allowed bound states, resulting in an energy spectrum that is finite, discrete, and bounded. Furthermore, the domain in \(k\)-space where these stationary states exist shrinks as the number of excited states increases. 

Both the probability density and the current density exhibit explicit dependence on the valley and band indices. In particular, due to the tilt of the Dirac cones, the probability current remains nonzero even in the ground state—behavior that distinguishes these systems from ideal graphene. Importantly, the method developed in this work enables a direct solution to the eigenvalue problem without resorting to supersymmetric quantum mechanics, Lorentz or similarity (non-unitary) transformations, thereby offering a more transparent alternative to previously established approaches\cite{Ateş_2023, Kuru2009, campa2024, Mojica-Zárate_2023, Lukose2007, O-Campa_2024}.

Despite the methodological differences, our results for the first two cases are fully consistent with previous findings obtained. Furthermore, all cases successfully reproduce the known spectrum of graphene under magnetic fields in the appropriate limit, thus validating our theoretical framework and confirming its broader applicability.

Finally, we emphasize that although polynomial solutions to Heun equations are known, they are typically expressed as power series whose coefficients satisfy the so-called three-term recurrence relation \cite{10.1063/1.4893997,MELIKDZHANIAN2021125037}. However, in the present work, we have not only found such polynomial solutions but also managed to express them in terms of well-known families of orthogonal polynomials.

\textbf{Acknowledgments}
The authors acknowledge financial support from CONAHCYT Project FORDECYT-PRONACES/61533/2020. EDB also acknowledges the SIP-IPN research grant 20254000. DOC gratefully acknowledges the financial support from SECIHTI through the postdoctoral fellowship, CVU number 712322.

\textbf{Data Availability Statement} All data generated or analyzed during this study are included in this published article.

\bibliographystyle{unsrt}
\bibliography{biblio}
------------------------
\end{document}